\documentclass[aip,sd,amsmath,amssymb,numerical]{revtex4-1}
\usepackage{graphicx}
\usepackage{dcolumn}
\usepackage{bm}

\begin{document}

\preprint{AIP/123-QED}


\title{Understanding Nucleic Acid Structural Changes by Comparing Wide-Angle X-ray Scattering (WAXS) Experiments to Molecular Dynamics Simulations} 




\author{Suzette A. Pabit}
\author{Andrea M. Katz}
\affiliation{School of Applied and Engineering Physics, Cornell University, Ithaca, NY 14853, USA}

\author{Igor S. Tolokh}
\affiliation{Department of Computer Science, Virgina Tech, Blacksburg, VA 24061, USA}
\author{Aleksander Drozdetski}
\affiliation{Department of Physics, Virgina Tech, Blacksburg, VA 24061, USA}

\author{Nathan Baker}
\affiliation{Pacific Northwest National Laboratory, Richland, WA 99352, USA}

\author{Alexey V. Onufriev}
\affiliation{Department of Computer Science, Virgina Tech, Blacksburg, VA 24061, USA}
\affiliation{Department of Physics, Virgina Tech, Blacksburg, VA 24061, USA}

\author{Lois Pollack}
\affiliation{School of Applied and Engineering Physics, Cornell University, Ithaca, NY 14853, USA}


\date{\today}

\begin{abstract}
Wide-angle x-ray scattering (WAXS) is emerging as a powerful tool for increasing the resolution of solution structure measurements of biomolecules. Compared to its better known complement, small angle x-ray scattering (SAXS), WAXS targets higher scattering angles and can enhance structural studies of molecules by accessing finer details of solution structures. Although the extension from SAXS to WAXS is easy to implement experimentally, the computational tools required to fully harness the power of WAXS are still under development. Currently, WAXS is employed to study structural changes and ligand binding in proteins; however the methods are not as fully developed for nucleic acids. Here, we show how  WAXS can qualitatively characterize nucleic acid structures as well as the small but significant structural changes driven by the addition of multivalent ions. We show the potential of WAXS to test all-atom molecular dynamics (MD) simulations and to provide insight in understanding how the trivalent ion cobalt(III) hexammine (CoHex) affects the structure of RNA and DNA helices. We find that MD simulations capture the RNA structural change that occurs due to addition of CoHex. 
\end{abstract}


\maketitle 

\section{Introduction}

The accessibility of synchrotron radiation sources for Small Angle X-ray Scattering (SAXS) experiments has enabled new methods for measuring low resolution structures of biological macromolecules \cite{Perez2012,Skou2014}. The spatial resolution $d$ of a SAXS measurement is largely determined by the highest $q$ value of the measurement, given by $d = 2\pi/q$, where $q$ is the momentum transfer, $q = (4\pi/\lambda) \cdot \sin(2\theta/2)$, ~$\lambda$ is the x-ray wavelength and $2\theta$ is the scattering angle\cite{Makowski2010}. Extension from small-angle to wide-angle x-ray scattering (WAXS) is appealing because scattering at wider angles yields higher resolution structural features\cite{Graewert2013,Makowski2010}. For example, even a modest maximum $q$ at around 1.0 {\AA}$^{-1}$ corresponds to features that are in the $5-10$ {\AA} spatial range. Implementation of WAXS is simple and often requires a trivial modification to a SAXS beamline: the detector is moved closer to the sample to capture x-rays scattered to higher angles. It is even possible to simultaneously acquire SAXS and WAXS data by placing a detection window near the sample\cite{Allaire2011}. WAXS provides information about length scales of critical structural importance, and is particularly relevant for studies of nucleic acid structures. For example, in DNA, peaks in the WAXS regime of $0.4 < q < 1.0$ {\AA}$^{-1}$ arise from interstrand pair distance correlations from major and minor groove spacing and helix radius\cite{Zuo2006}.

Despite the relative ease of acquiring WAXS data, the challenge in its application arises from the interpretation of the measured scattering profiles. On these shorter length scales, for example, the contribution of the solvent, including the hard-to-model hydration shell around biomolecules, becomes significant\cite{Park2009}. A number of experimental and computational tools for the analysis of WAXS structures of biomolecules are available but all rely on the availability of high resolution atomic coordinates for comparison\cite{Makowski2010, Bardhan2009, Knight2015}. Since atomic coordinates are not readily available for many biomolecules, this limits the applicability of WAXS analysis software. Moreover, most of these analysis tools are geared towards studies of protein structural fluctuations and protein ligand binding \cite{Makowski2010,Virtanen2011,Minh2013}, where the hydration layer is reasonably well described. In contrast, because of the highly charged nature of nucleic acids, accurate models of the dense solvent layer adjacent to the molecule are still under development. A method to calculate nucleic acid SAXS and WAXS profiles that includes not only the hydration shell, but also charge compensating counterions has recently been introduced\cite{Nguyen2014} but still requires input of matching atomic coordinates of nucleic acids structures.   

The goal of this paper is to extend the use of WAXS to studies of nucleic acids. We discuss two cases where interpretation of WAXS experimental data is possible.  First, when key structural features are present in the nucleic acid scattering profiles, we can compare the location of the peaks and valleys to those calculated from existing structural models of nucleic acids based on canonical helices and Molecular Dynamics (MD) simulation results. In this way, WAXS data can provide experimental benchmarks that can lead to development and refinement of MD simulations. When data and models agree, MD results can help visualize WAXS-resolved structural features. The  second way to interpret nucleic acid WAXS data is to look at changes in the WAXS profiles induced by the addition of ligands or multivalent ions. Intensity difference curves have been used in the study of WAXS structures in proteins, particularly for comparing different time points in time-resolved experiments\cite{Cammarata2008,Arnlund2014}. 
Here, we show that  a comparison of intensity difference curves (MD to experiment) is a useful way to interpret changes in the nucleic acid structure on the nanometer length scale. 

\section{Experimental and Computational Details}

\subsection{Background on Wide-Angle X-ray Scattering (WAXS)}

Small Angle X-ray Scattering (SAXS) is a useful experimental technique that typically provides low resolution structural information: a measure of the radius of gyration, pair-distance distribution function and computation of shape envelopes of proteins and nucleic acids in solution \cite{Perez2012,Skou2014}. A standard SAXS set-up is shown schematically in the top panel of Figure 1. X-rays are incident on a biological sample, the scattered x-rays pass through a 1 to 2 m vacuum flight path and are collected by a photon detector. The collected scattered intensity $I$ is displayed as a function momentum transfer $q$ given by $q = (4\pi/\lambda) \cdot \sin(2\theta/2)$, ~$\lambda$ is the x-ray wavelength and $2\theta$ is the scattering angle\cite{Makowski2010}. 

\begin{figure}[h]
\centering
\caption{Schematic of solution-based x-ray scattering experiments in the small-angle (top) and wide-angle (bottom) regimes. The main difference between SAXS and WAXS  set-ups is the sample-to-detector distance. The wider angles in WAXS experiments allow access to high resolution information in the $5-10$ {\AA} spatial range.} 
\end{figure}

Wide-angle x-ray scattering (WAXS) experiments, shown schematically in the bottom panel of Figure 1, typically use the same set-up as SAXS but with a shorter scattering flight path which allow access to higher resolution structural information\cite{Graewert2013,Makowski2010}. The spatial resolution $d$ of a scattering experiment is largely determined by the highest $q$ value of the measurement, given by $d = 2\pi/q$. For example, shortening the sample-to-detector distance to $0.4-0.5$ m allows access to a higher $q$-range, $q$ at around 1.0 {\AA}$^{-1}$ corresponds to features that are in the $5-10$ {\AA} spatial range. WAXS data acquisition is becoming a popular addition to SAXS experiments by simple installation of a WAXS detection window closer to the sample\cite{Allaire2011}. Therefore, there is need for development of more analytical tools to understand WAXS data.

\subsection{Sample Preparation and Experimental Conditions}

Double-stranded 25 base-pair (bp) DNA and RNA constructs were annealed from single-stranded nucleic acids purchased from IDT (Coralville, IA). We used the same mixed sequence described in our previous publications, GCA TCT GGG CTA TAA AAG GGC GTC G (U replacing T for the RNA strands)
\cite{Li2011,Tolokh2014}. Cobalt(III) Hexammine chloride, Co(NH$_{3}$)$_{6}$Cl$_{3}$ (CoHex), sodium chloride, NaCl, and the buffers used were purchased from Sigma-Aldrich (St. Louis, MO). The duplexes were extensively dialyzed using Amicon Ultra Concentrators (Millipore, Billerica, MA) in pH 7.0 Na-MOPS buffer containing either 100 mM NaCl only or 100 mM NaCl and 0.8 mM Co(NH$_{3}$)$_{6}$Cl$_{3}$. The monovalent salt concentrations were chosen to prevent trivalent CoHex induced precipitation and focus the study on CoHex-nucleic acid interactions in the pre-condensed solution phase. The WAXS experiments were carried out at the G1 station at the Cornell High Energy Synchrotron Source (CHESS). Using a scattering flight path with a 0.455 m sample to detector distance, we reached a $q_{max} = 0.95$ {\AA}$^{-1}$, which is sensitive to correlation lengths of $d_{min} = 2\pi/ q_{max} = 6.6$ {\AA}. This allows us to target the WAXS region of $0.4 < q < 0.95$ {\AA}$^{-1}$ which corresponds to the length scales of the helix radius and the minor and major groove spacing\cite{Zuo2006}. We purposely avoided $q > 1.0$ {\AA}$^{-1}$ where contributions from solvent scattering become more intense \cite{Makowski2010}. The x-ray energy used was 10.60 keV and scattered x-rays were imaged using a low-noise photon counting area detector (Pilatus 100K, Dectris, Baden, Switzerland). Duplex concentrations were about 450 $\mu$M. The samples, with volumes of about 30 $\mu$L, were kept in a 2-mm quartz capillary (Hampton Research, Aliso Viejo, CA) and radiation damage was prevented by oscillating the sample during the data collection. Signals from the buffer background were subtracted from the data and absolute calibration was performed using water as a calibrant as described previously\cite{Orthaber2000,Nguyen2014}. 
Data analysis was done using code written in MATLAB (Mathworks, Natick, MA). 
A more detailed description of the experimental measurements are provided in Ref. \cite{Nguyen2014}.

\subsection{Molecular Dynamamics (MD) Simulations and Comparison with WAXS Data}

\begin{figure}[h]
\centering
\caption{ DNA and RNA double helices have the same charge, $-2e$ per base pair, but different helical structures. Shown here are 25-base-pair long canonical double helices of B-form DNA (left) and A-form RNA (right) generated by Nucleic Acid Builder (NAB)\cite{Macke1998} using the sequences used in the experiment. These helices are the launching point of the Molecular Dynamics simulations used and the basis for comparison of the WAXS experimental spectra to the canonical spectra generated using the program CRYSOL\cite{Svergun1995}.  The canonical structures shown are drawn using Acelerys Discovery Studio program suite.} 
\end{figure}

MD simulations of RNA and DNA constructs in explicit water and monovalent ions (NaCl) were carried out in the presence and absence of CoHex ions using AMBER 12 \cite{Case2005} and ff99bsc0 force fields \cite{Cheatham1999,Perez2007} which includes monovalent ion parameters\cite{Aqvist1990, Smith1994}. All simulated systems included about 16880 TIP3P water molecules\cite{Jorgensen1983}. The systems with CoHex included 16 CoHex ions using parameters from Cheatham and Kollman\cite{Cheatham1997} 
which are part of the AMBER12 distribution. The systems without CoHex were neutralized by 48 Na$^{+}$ ions.  Monovalent salt background in all the systems was accounted for by additional 24 NaCl ion pairs. The starting structures for the MD simulations were generated from canonical helices (A-form for RNA and B-form for DNA) built using Nucleic Acid Builder (NAB)\cite{Macke1998} and shown in Figure 2. All the systems were initially equilibrated for 0.5 ns in canonical ensemble (NVT) and 0.5 ns in isothermal-isobaric ensemble (NPT) using 1 fs time step and achieving 1 atm pressure and 300 K temperature. The following simulations were carried out in NVT ensemble using 2 fs time step, periodic boundary conditions, the particle mesh Ewald method at 300 K maintained using Langevin dynamics with the collision frequency of 1 ps$^{-1}$. In the first 140 ns of  simulation, a restraining harmonic potential with a force constant of 50 kcal/mol/{\AA}$^{2}$ was applied to the canonical forms of the RNA and DNA molecules. The restraining force was removed and the simulations were allowed to proceed for another 100-200 ns. 

To compare the MD results with the WAXS data, we extracted 100-500 snapshots\cite{Chen2014} from each trajectory in PDB format and generated WAXS profiles using the program CRYSOL\cite{Svergun1995}.  CRYSOL is a commonly-used program for calculating the solution scattering profiles of macromolecules with known atomic coordinates. 
Though other WAXS solvers are available and a simple test of the program WAXSiS\cite{Knight2015} yielded similar profiles, we chose CRYSOL for speed of the calculations and ease of handling hundreds of PDB snapshots. The CRYSOL input settings were configured so that the output scattering profiles are fitted to the experimental data. All other parameters were set to the CRYSOL default values (e.g. solvent density set to pure water). We report all CRYSOL results in absolute intensity units ($e^{2}$). To focus on the nucleic acid structural features, the ions were generally not included in the CRYSOL calculations. The inclusion of the ions is discussed in the latter part of the paper (See Discussion Section IV. A.). Profiles generated from hundreds of PDB snapshots were pooled together into two groups, MD profiles with the restraining force in place (``restrained'' profiles) and MD profiles after the restraining force was released (``free'' profiles). The pools were averaged and these averages were compared to the data. 

\section{WAXS Results}

\subsection{Nucleic Acids in NaCl}

\begin{figure*}
\centering
\caption{Comparison of wide-angle scattering data from (A) DNA and (B) RNA in 100 mM NaCl to canonical helices built using Nucleic Acid Builder (NAB) and structural ensembles generated by  ``restrained'' MD simulations. The small differences between NAB and ``restrained'' MD reflects the small deviation from the restraining harmonic potential.  A similar comparison is made between (C) DNA and (D) RNA data and ``free'' MD structures after the restraining force is removed. We magnified the features in the WAXS regime, $q > 0.4$ {\AA}$^{-1}$, and show the full scattering profile as insets.} 
\end{figure*}

Figure 3 compares background subtracted experimental WAXS profiles of nucleic acids in NaCl to canonical helices and MD simulation profiles. We show intensity, $I$ vs.  $q$ curves in logarithmic scale to enhance structural features in the WAXS regime, the higher-$q$ region of the dataset ($q > 0.4$ {\AA}$^{-1}$). 
Displaying the data in this form allows us to easily assess whether the general features and peaks of the experimental profiles agree or disagree with the simulations and the canonical helices built using NAB. In Figure 3A, we compare our experimental profiles of DNA in NaCl to a canonical B-form helix and MD simulations in NaCl and explicit water with the restraining force in place (``restrained'' profiles). Since the ``restrained'' MD models are derived from NAB, the calculated profiles are very similar with very small deviations due to the restraining harmonic potential. Note the relative agreement between the experiment and the calculated ``restrained'' MD and NAB profiles, exhibiting the characteristic features that were seen in previous measurements\cite{Zuo2006} on B-form DNA with a peak measured at $q = 0.48$ {\AA}$^{-1}$ and a shallow peak detected at $q = 0.72$ {\AA}$^{-1}$. Although agreement is qualitative, the peak positions are in the right place. Thus, scattering profiles for DNA in NaCl compare favorably with predictions of ``restrained'' MD simulations. In contrast, the scattering profile of RNA in NaCl shown in Figure 3B looks substantially different from the ``restrained'' MD profiles. As expected, the MD simulations results using ``restrained'' RNA recapitulate the features exhibited in the NAB canonical A-form structure  because of the restraining potential. However, both curves disagree with the experimental data. The first peak in the experimental WAXS profiles for RNA is too shallow (more like a shoulder than a peak) and shifted to the left compared to the ``restrained'' MD profile, the second peak is also shifted to the left.

In Figures 3C and 3D, we compare the experimental data with the MD simulation results after the restraining force has been released (``free'' profiles). Interestingly, the WAXS profiles computed for ``free'' DNA, Figure 3C, no longer capture the agreement with experiment as seen in the ``restrained'' models. A close examination of the WAXS profiles around $q = 0.5$ {\AA}$^{-1}$ and $q = 0.7$ {\AA}$^{-1}$ of unrestrained DNA curves in Figure 3C show that the peak positions are completely different and the simulation curves appear ``out-of-phase'' with the data. On the other hand, the features of the RNA WAXS data in NaCl are better captured by the MD simulations after the restraining force has been released. In Figure 3D, we note that the shoulder in the ``free'' MD near $q = 0.4$ {\AA}$^{-1}$ also appears in the experimental curves. However, the second peak position is not exactly in the same place. Nonetheless, at $q = 0.71$ {\AA}$^{-1}$, it is closer to the experimental peak at $q = 0.74$ {\AA}$^{-1}$ than the peak value of $q = 0.82$ {\AA}$^{-1}$ for the ``restrained'' MD. The peak positions in Figures 3C and 3D show that there is definite improvement in the simulated RNA spectra when unrestrained MD is performed but not in DNA. This result suggests that in an NaCl solution, duplex RNA does not have a rigid canonical A-form structure while DNA maintains the B-form canonical structure in NaCl (Figure 3A).

\subsection{Nucleic Acids in CoHex}

\begin{figure*}
\centering
\caption{Comparison of wide-angle scattering data from nucleic acids in 0.8 mM CoHex and 100 mM NaCl solution to ``restrained'' and ``free'' MD simulations. (A) and (C) are for DNA and (B) and (D) are for RNA. The number of CoHex and NaCl in the simulation box complements the experiment. We put emphasis in the features in the WAXS regime, $q > 0.4$ {\AA}$^{-1}$, by magnification and show the full scattering profile as insets.}
\end{figure*}

In Figure 4, we examine structural changes in DNA and RNA induced by the addition of multivalent ions.  Cobalt(III) Hexammine, [Co(NH$_{3}$)$_{6}$]$^{3+}$ or CoHex, is a trivalent ion that is widely used to facilitate and study DNA condensation\cite{Widom1980,Pelta1996}. This highly charged spherical ion has a dramatically different effect on short double helices made of DNA and RNA. The addition of a small amount of CoHex ions precipitates short DNA helices from solution, while RNA molecules of similar sequence remain soluble\cite{Li2011,Tolokh2014}. The efficiency of condensation appears to be primarily related to the mode of CoHex binding, which is mostly determined by the helix geometry, although other factors like nucleotide sequence do contribute. 

In a related study\cite{Tolokh2014}, we used UV spectroscopy and circular dichroism (CD) to monitor CoHex-induced precipitation of DNA and RNA helices, and compared MD simulations of the different duplexes. We showed that simulations of B-form DNA suggest that CoHex ions decorate the outer surface of the structure, while MD simulations of A-like RNA suggest that the ions are drawn deep into the negatively charged major groove. There, we discussed how ion placement influences precipitation. However, a limitation of that previous study \cite{Tolokh2014} lies in our inability to directly compare MD simulations to experimental data. Although CD reports spectral changes in RNA, there is no straightforward way to correlate the changes in nucleic acid CD spectra with MD predictions. In fact, our attempts to use the CD modeling software Dichrocalc \cite{Bulheller2009}, which was written for proteins, yielded results that substantially differ from the CD data. Here, we explore the use of WAXS as an experimental way to check if changes in the structures of DNA and RNA duplexes due to CoHex are predicted by simulation. 

Although the signal from the CoHex ions is too small to be detected experimentally in the WAXS profiles, CoHex-driven changes in local structure, suggested by the CD data, should be noticeable. Figure 4 shows WAXS profiles of the nucleic acids in a CoHex-NaCl solution and compares the profiles with simulations. Although we observe some  lineshape broadening in the WAXS peaks of the DNA samples, the peak positions do not move after addition of CoHex. The same B-form DNA structure is still captured in Figure 4A with ``restrained'' MD. This behavior is expected from DNA molecules since Circular Dichroism (CD) measurements showed that CD changes in DNA with and without CoHex are minimal while a large spectral shift in the CD spectra for RNA was observed upon addition of CoHex \cite{Tolokh2014}. In agreement with CD, distinct changes are observed in the RNA WAXS profiles upon addition of CoHex. In fact, in the presence of CoHex, the WAXS structures of RNA appear more like canonical A-form helix and have peaks in similar positions to the ``restrained'' MD (Figure 4B). Similar to DNA in NaCl, the ``free'' MD profiles for DNA in CoHex show disagreement in peak positions (Figure 4C). For RNA, peak positions in the ``free'' MD profiles qualitatively agree with the data (Figure 4D).

\subsection{Intensity Difference Profiles Before and After CoHex}

\begin{figure*} [t!]
\centering
\caption{Intensity difference profiles, $\log(I_\text{NaCl})-\log(I_\text{CoHex})$, comparing nucleic acid experimental data to MD results when the molecules are ``restrained'' by a harmonic potential (A and B) and when molecules are ``free'' (C and D). The figure legends shown in panel (C) for DNA also apply to panel (D) for RNA. The best qualitative agreement with experiment happens when RNA is unrestrained in the MD simulations.}
\end{figure*}

To better compare changes in structural features in WAXS profiles due to addition of CoHex, we look at differences in the scattering intensity following the method used in Refs. \cite{Makowski2010,Cammarata2008,Arnlund2014,Fischetti2004}. In Figure 5, we show experimental intensity difference profiles, $\log(I_\text{NaCl})-\log(I_\text{CoHex})$, and compare them to the difference profiles computed from the simulated curves.  Since all the experimental data shown have absolute calibration and the MD profiles calculated using CRYSOL are made to fit the data, no scaling was applied prior to the subtraction. Looking at all panels in Figure 5, we immediately see that the key features in the intensity difference curves are best captured by MD results shown in Figure 5D where unrestrained or ``free'' MD difference profiles are compared to the experimental difference profiles for RNA. The difference peaks at $q = 0.3$ {\AA}$^{-1}$ and at $q = 0.7$ {\AA}$^{-1}$ in the RNA experimental curves also appear in the MD simulations albeit with different amplitudes. Makowski and coworkers\cite{Makowski2010,Fischetti2004} have shown that when intensity difference profiles in WAXS are calculated using CRYSOL\cite{Svergun1995}, the difference curves correspond more closely to experimental results than the absolute intensities. This result is quite interesting: it suggests that CoHex-induced changes in the RNA structure are well-represented by the ``free'' MD simulations after the restraining force on RNA has been removed.

MD models of DNA are harder to describe because features in the difference profiles are not as distinct as they are for RNA. However, even with less pronounced features, we find that the DNA data are more consistent with MD results from ``restrained'' DNA, not ``free'' DNA. The ``restrained'' MD-derived difference curve is largely featureless  (Figure 5A) while the `free'' MD-derived difference curve  (Figure 5C) has peaks and valleys whose locations are inconsistent with the data. Figure 5B for ``restrained'' RNA also shows disagreement in the position of peaks and valleys. Thus, we find that comparison of intensity difference curves to models is not only a useful way to denote changes in the nucleic acid structure, but is also a robust method to identify good models and discard models that do not reflect experimental data. The agreement in Figure 5D between MD and experiment is still qualitative but the disagreement in Figure 5B and 5C is quite notable.

\section{Discussion}

\subsection{RNA structure changes upon addition of Cohex} 

\begin{figure*} [t]
\centering
\caption{WAXS profiles comparing RNA data to ``free'' MD and the canonical A-form helix from NAB in (A) NaCl only and (B) NaCl/CoHex solution. Insets show representative structures of RNA from MD snapshots. Double-stranded RNA with CoHex appears to have a shorter end-to-end distance and a WAXS profile that has features more similar to the canonical A-form helix. }
\end{figure*}

These results can now be used to gain physical insight into the origin of the structural changes.  The need to use unrestrained ``free'' MD simulations for RNA suggests that CoHex introduces structural changes to the RNA  helix. In Figure 6, we show experimental WAXS profiles, $I$ vs. $q$, in logarithmic scale, comparing RNA to ``free'' MD results and the canonical A-form helix from NAB. Without CoHex, Figure 6A shows RNA experimental profiles favoring the ``free'' MD profiles with the shoulder around $q = 0.45$ {\AA}$^{-1}$. However, the shallow peak seen in the data at $q = 0.74$ {\AA}$^{-1}$ is shifted to the left in the MD simulations (peak at $q = 0.71$ {\AA}$^{-1}$). In the presence of CoHex (Figure 6B), both RNA experimental curves and the results from the ``free'' MD appear to have features more similar to the canonical A-form profile. When the ions are turned off in CRYSOL (our default setting), the second peak around $q = 0.81$ {\AA}$^{-1}$ becomes more pronounced in the ``free'' MD profile with CoHex, which implies that CoHex changes the actual structure of the RNA, not just the counterion cloud or the hydration layer. The smearing of the $q = 0.81$ {\AA}$^{-1}$ peak when ions are turned on suggests diffuse positioning of the CoHex ions in the vicinity of $d = 2\pi/q = 7.8$ {\AA}, consistent with our picture of CoHex ions deep in the A-form major groove\cite{Tolokh2014}. Figure 6 insets show representative snapshots of the RNA structure from the unrestrained MD simulation with and without CoHex. From the pool of MD-derived PDB snapshots, we measured the average distance between the 3'-end phosphates of the base-paired strands. The end-to-end phosphate distance for the RNA in NaCl was 75 $\pm$ 3 {\AA} vs. 66 $\pm$ 2 {\AA} for the RNA in CoHex/NaCl suggesting a shortening of the structure in the presence of CoHex.

\subsection{Current MD Force-fields work better for RNA}

Here, we investigate the source of the disagreement between WAXS experiments and the unrestrained MD simulations of DNA. We compare the structural parameters of canonical A-form and B-form helices to the MD trajectories of the DNA and RNA sequences used in the study. Table 1 shows parameters from the first 100 ns following the release of the positional restraints and relaxation of the free structure. This corresponds to simulation time of 160--260 ns, the time scale consistent with CRYSOL-generated WAXS profiles averaged in Figures 3 and 4. We also show the parameters of a DNA dodecamer (DD) before and after simulations are performed. These parameters are also displayed in Table 1 below.

\begin{table}[h!]
\begin{center}
    \begin{tabular}{ | l | l | l | l |}
    \hline
    System & X-disp {\AA} & Rise {\AA} & Twist (deg) \\ \hline \hline
    Canonical A-form helix & -4.73 & 2.56 & 32.70 \\ \hline
    RNA in NaCl & -4.94   & 2.72 &	31.11\\ \hline
    RNA in NaCl/CoHex & -4.54 & 2.42  & 33.64 \\ \hline
    Canonical B-form helix & 0.11 & 3.38 & 36.00 \\ \hline
    DNA in NaCl & -2.39  &	3.14 &	32.63  \\ \hline
    DNA in NaCl/CoHex & -2.32 & 3.08 & 33.11  \\ \hline
    DD (PDB: 1BNA) & 0.13 & 3.35 & 36.09 \\ \hline
    DD in NaCl for 100 ns & -1.37 & 3.22 & 34.17 \\ \hline
    \end{tabular}
\end{center}
\caption{Structural parameters for the canonical and simulated helices and the B-DNA dodecamer (DD, PDB ID 1BNA). The simulated DNA and RNA are the 25-bp fragments with the same sequence as those used in the experiment, simulated with and without CoHex counterions. The DD in NaCl was simulated unrestrained for 100 ns using the same MD simulations conditions described in the main text, and the entire trajectory analyzed. To extract structural parameters, cpptraj tool in the AmberTools package was used, with the average values reported in the table. }
\end{table}

In Table 2, we show the same parameters for DNA and RNA after a longer simulation set, up to an additional 280 ns. The parameters appear to be quite stable during the entire length of the trajectories, and there is no significant difference between the time frame used for CRYSOL snapshots (Table 1) and the next 100 ns. 

\begin{table}[h!]
\begin{center}
    \begin{tabular}{ | l | l | l | l |}
    \hline
    System & X-disp {\AA} & Rise {\AA} & Twist (deg) \\ \hline \hline
    Canonical A-form helix & -4.73 & 2.56 & 32.70 \\ \hline
    RNA in NaCl & -4.90  &	2.74 &	31.08 \\ \hline
    RNA in NaCl/CoHex & -4.64 & 2.40 & 33.40 \\ \hline
    Canonical B-form helix & 0.11 & 3.38 & 36.00 \\ \hline
    DNA in NaCl & -2.40  &	3.15 &	32.59 \\ \hline
    DNA in NaCl/CoHex & -2.35 & 3.08 & 33.02 \\ \hline   
 \end{tabular}
\end{center}
\caption{Structural parameters for the simulated unrestrained DNA and RNA analyzed with a longer trajectory, up to an additional 280 ns. The parameters for A-form and B-form helices are also shown for comparison.}
\end{table}

These tables show convergence of MD simulations. The helical parameters reach equilibrium for the snapshots used for CRYSOL. The tables also show that on time scales of hundreds of nanoseconds, the structural parameters of unrestained simulated mixed DNA duplexes drift away from the canonical B-form helix. 
The canonical parameters were originally reported by Arnott and Hukins\cite{Arnott1972}. The simulated structural parameters of the mixed sequence 25bp DNA (expected to be in B-form) and DNA dodecamer (DD) both deviate significantly from the canonical parameters of B-form DNA, and appear to drift toward A-form parameters, with the deviation from canonical B-form increasing (significantly for X-displacement, and twist) for the longer fragment. This deviation for DD is consistent with simulation results reported previously\cite{Perez2008} and implies that the force field used in this study is not ideal for reproducing subtle features of B-form DNA. Future work should look at the effects of other force fields on DNA and RNA helices and investigate the effects of dynamical motions such as base-pair fraying at the ends of the helices when comparing MD simulations to WAXS data.  

At the same time, the addition of CoHex produces no significant effect (within the above set of parameters) on the simulated mixed sequence DNA structure. Thus, while the force-field may be biased such that canonical B-form of DNA is not the preferred structure for the mixed 25-bp fragments (which would explain the discrepancy between unrestrained simulations and experimental wide-angle scattering data), it still appears to reproduce ion-DNA interaction reasonably well, as evidenced by minimal change in DNA structure upon addition of CoHex, which is in agreement with CD\cite{Tolokh2014} and WAXS measurements. Interestingly, the decrease in RNA rise with the addition of CoHex, from 2.74 \AA~to 2.40 \AA~corresponds to the decrease in the end-to-end distance (75 \AA~to 66 \AA) and the change in both CD\cite{Tolokh2014} and WAXS spectra.  

The limitations of MD force-fields for DNA have also been discussed in a paper by Tiede and coworkers where they benchmarked the WAXS profiles to MD solutions and found a need to develop experimentally validated, supramolecular force fields\cite{Zuo2006}. Our findings reiterate theirs, and further the field by reporting results with RNA. What is surprising here is the better agreement between the unrestrained MD and the experimental data in RNA. The disagreement between our DNA WAXS data and the structures from the ``free'' MD for DNA arise from deviations of key structural features in the MD-generated ensemble from the values expected for canonical B-form DNA. Force field modification is beyond the scope of this work, however, we illustrate the use of WAXS as an experimental test. While the WAXS curves are not exactly predicted (on an absolute scale) by the force field, they present qualitative features that accurately reflect structural features of these duplexes, hence provide a useful experimental metric. 

\section{Conclusions}

In summary, we compared MD-derived structures to experimental WAXS data and showed the power of WAXS in investigating nucleic acid structural features and structural changes.  Although our analysis of WAXS data is currently limited by the need to compare to accessible PDB structures, comparison of intensity difference profiles from experimental data and those generated from MD-derived models provides a test of MD simulation predictions. Changes in RNA WAXS data with and without CoHex ions were well represented by changes in profiles computed from structures generated in unrestrained MD simulations. Changes in experimental DNA WAXS profiles are smaller and give confidence to the previous assumption \cite{Tolokh2014} to restrain DNA structures during MD simulations; restrained DNA is best used in situations when the study focuses on DNA surroundings, e.g. distribution of solvent components such as water or ions. Our study here showed that there are limitations in using unrestrained MD for DNA in comparison to WAXS profiles. However, even with the limitations shown, we were able to gain insight into molecular conformations by comparing unrestrained MD of RNA to experimental WAXS data. We demonstrate that CoHex affects RNA helices by shortening the end-to-end distance and forcing the molecule to adapt a more A-like conformation. The future application of computational methods that allow MD simulations to be guided by SAXS and WAXS data, a process that is currently being developed for proteins\cite{Chen2015}, holds great potential for studies of  nucleic acids.

\begin{acknowledgments}
We acknowledge funding support from NIH grant R01 GM-099450 and, in part, by NSF grant CNS-0960081 and the HokieSpeed supercomputer at Virginia Tech. A.M.K. was supported by the NSF Graduate Research Fellowship grant DGE-1144153. We thank Arthur Woll, Steve Meisburger, Huimin Chen and Pollack Lab Members for experimental assistance. WAXS data were acquired at the Cornell High Energy Synchrotron Source (CHESS). CHESS is supported by the NSF \& NIH/NIGMS via NSF award DMR-1332208, and the MacCHESS resource is supported by NIGMS award GM-103485. 
\\
\end{acknowledgments}

\nocite{*}

\bibliography{cohexwaxs_cited3}

\begin{thebibliography}{34}%
\makeatletter
\providecommand \@ifxundefined [1]{%
 \@ifx{#1\undefined}
}%
\providecommand \@ifnum [1]{%
 \ifnum #1\expandafter \@firstoftwo
 \else \expandafter \@secondoftwo
 \fi
}%
\providecommand \@ifx [1]{%
 \ifx #1\expandafter \@firstoftwo
 \else \expandafter \@secondoftwo
 \fi
}%
\providecommand \natexlab [1]{#1}%
\providecommand \enquote  [1]{``#1''}%
\providecommand \bibnamefont  [1]{#1}%
\providecommand \bibfnamefont [1]{#1}%
\providecommand \citenamefont [1]{#1}%
\providecommand \href@noop [0]{\@secondoftwo}%
\providecommand \href [0]{\begingroup \@sanitize@url \@href}%
\providecommand \@href[1]{\@@startlink{#1}\@@href}%
\providecommand \@@href[1]{\endgroup#1\@@endlink}%
\providecommand \@sanitize@url [0]{\catcode `\\12\catcode `\$12\catcode
  `\&12\catcode `\#12\catcode `\^12\catcode `\_12\catcode `\%12\relax}%
\providecommand \@@startlink[1]{}%
\providecommand \@@endlink[0]{}%
\providecommand \url  [0]{\begingroup\@sanitize@url \@url }%
\providecommand \@url [1]{\endgroup\@href {#1}{\urlprefix }}%
\providecommand \urlprefix  [0]{URL }%
\providecommand \Eprint [0]{\href }%
\providecommand \doibase [0]{http://dx.doi.org/}%
\providecommand \selectlanguage [0]{\@gobble}%
\providecommand \bibinfo  [0]{\@secondoftwo}%
\providecommand \bibfield  [0]{\@secondoftwo}%
\providecommand \translation [1]{[#1]}%
\providecommand \BibitemOpen [0]{}%
\providecommand \bibitemStop [0]{}%
\providecommand \bibitemNoStop [0]{.\EOS\space}%
\providecommand \EOS [0]{\spacefactor3000\relax}%
\providecommand \BibitemShut  [1]{\csname bibitem#1\endcsname}%
\let\auto@bib@innerbib\@empty
\bibitem [{\citenamefont {P\'{e}rez}\ and\ \citenamefont
  {Nishino}(2012)}]{Perez2012}%
  \BibitemOpen
  \bibfield  {author} {\bibinfo {author} {\bibfnamefont {J.}~\bibnamefont
  {P\'{e}rez}}\ and\ \bibinfo {author} {\bibfnamefont {Y.}~\bibnamefont
  {Nishino}},\ }\href@noop {} {\bibfield  {journal} {\bibinfo  {journal}
  {Current Opinion in Structural Biology}\ }\textbf {\bibinfo {volume} {22}},\
  \bibinfo {pages} {670--678} (\bibinfo {year} {2012})}\BibitemShut {NoStop}%
\bibitem [{\citenamefont {Skou}, \citenamefont {Gillilan},\ and\ \citenamefont
  {Ando}(2014)}]{Skou2014}%
  \BibitemOpen
  \bibfield  {author} {\bibinfo {author} {\bibfnamefont {S.}~\bibnamefont
  {Skou}}, \bibinfo {author} {\bibfnamefont {R.~E.}\ \bibnamefont {Gillilan}},
  \ and\ \bibinfo {author} {\bibfnamefont {N.}~\bibnamefont {Ando}},\
  }\href@noop {} {\bibfield  {journal} {\bibinfo  {journal} {Nature protocols}\
  }\textbf {\bibinfo {volume} {9}},\ \bibinfo {pages} {1727--39} (\bibinfo
  {year} {2014})}\BibitemShut {NoStop}%
\bibitem [{\citenamefont {Makowski}(2010)}]{Makowski2010}%
  \BibitemOpen
  \bibfield  {author} {\bibinfo {author} {\bibfnamefont {L.}~\bibnamefont
  {Makowski}},\ }\href@noop {} {\bibfield  {journal} {\bibinfo  {journal}
  {Journal of Structural and Functional Genomics}\ }\textbf {\bibinfo {volume}
  {11}},\ \bibinfo {pages} {9--19} (\bibinfo {year} {2010})}\BibitemShut
  {NoStop}%
\bibitem [{\citenamefont {Graewert}\ and\ \citenamefont
  {Svergun}(2013)}]{Graewert2013}%
  \BibitemOpen
  \bibfield  {author} {\bibinfo {author} {\bibfnamefont {M.~A.}\ \bibnamefont
  {Graewert}}\ and\ \bibinfo {author} {\bibfnamefont {D.~I.}\ \bibnamefont
  {Svergun}},\ }\href@noop {} {\bibfield  {journal} {\bibinfo  {journal}
  {Current Opinion in Structural Biology}\ }\textbf {\bibinfo {volume} {23}},\
  \bibinfo {pages} {748--754} (\bibinfo {year} {2013})}\BibitemShut {NoStop}%
\bibitem [{\citenamefont {Allaire}\ and\ \citenamefont
  {Yang}(2011)}]{Allaire2011}%
  \BibitemOpen
  \bibfield  {author} {\bibinfo {author} {\bibfnamefont {M.}~\bibnamefont
  {Allaire}}\ and\ \bibinfo {author} {\bibfnamefont {L.}~\bibnamefont {Yang}},\
  }\href@noop {} {\bibfield  {journal} {\bibinfo  {journal} {Journal of
  Synchrotron Radiation}\ }\textbf {\bibinfo {volume} {18}},\ \bibinfo {pages}
  {41--44} (\bibinfo {year} {2011})}\BibitemShut {NoStop}%
\bibitem [{\citenamefont {Zuo}\ \emph {et~al.}(2006)\citenamefont {Zuo},
  \citenamefont {Cui}, \citenamefont {Merz}, \citenamefont {Zhang},
  \citenamefont {Lewis},\ and\ \citenamefont {Tiede}}]{Zuo2006}%
  \BibitemOpen
  \bibfield  {author} {\bibinfo {author} {\bibfnamefont {X.}~\bibnamefont
  {Zuo}}, \bibinfo {author} {\bibfnamefont {G.}~\bibnamefont {Cui}}, \bibinfo
  {author} {\bibfnamefont {K.~M.}\ \bibnamefont {Merz}}, \bibinfo {author}
  {\bibfnamefont {L.}~\bibnamefont {Zhang}}, \bibinfo {author} {\bibfnamefont
  {F.~D.}\ \bibnamefont {Lewis}}, \ and\ \bibinfo {author} {\bibfnamefont
  {D.~M.}\ \bibnamefont {Tiede}},\ }\href@noop {} {\bibfield  {journal}
  {\bibinfo  {journal} {PNAS}\ }\textbf {\bibinfo {volume} {103}},\ \bibinfo
  {pages} {3534--9} (\bibinfo {year} {2006})}\BibitemShut {NoStop}%
\bibitem [{\citenamefont {Park}\ \emph {et~al.}(2009)\citenamefont {Park},
  \citenamefont {Bardhan}, \citenamefont {Roux},\ and\ \citenamefont
  {Makowski}}]{Park2009}%
  \BibitemOpen
  \bibfield  {author} {\bibinfo {author} {\bibfnamefont {S.}~\bibnamefont
  {Park}}, \bibinfo {author} {\bibfnamefont {J.~P.}\ \bibnamefont {Bardhan}},
  \bibinfo {author} {\bibfnamefont {B.}~\bibnamefont {Roux}}, \ and\ \bibinfo
  {author} {\bibfnamefont {L.}~\bibnamefont {Makowski}},\ }\href@noop {}
  {\bibfield  {journal} {\bibinfo  {journal} {Journal of Chemical Physics}\
  }\textbf {\bibinfo {volume} {130}},\ \bibinfo {pages} {134114} (\bibinfo
  {year} {2009})}\BibitemShut {NoStop}%
\bibitem [{\citenamefont {Bardhan}, \citenamefont {Park},\ and\ \citenamefont
  {Makowski}(2009)}]{Bardhan2009}%
  \BibitemOpen
  \bibfield  {author} {\bibinfo {author} {\bibfnamefont {J.}~\bibnamefont
  {Bardhan}}, \bibinfo {author} {\bibfnamefont {S.}~\bibnamefont {Park}}, \
  and\ \bibinfo {author} {\bibfnamefont {L.}~\bibnamefont {Makowski}},\
  }\href@noop {} {\bibfield  {journal} {\bibinfo  {journal} {Journal of Applied
  Crystallography}\ }\textbf {\bibinfo {volume} {42}},\ \bibinfo {pages}
  {932--943} (\bibinfo {year} {2009})}\BibitemShut {NoStop}%
\bibitem [{\citenamefont {Knight}\ and\ \citenamefont
  {Hub}(2015)}]{Knight2015}%
  \BibitemOpen
  \bibfield  {author} {\bibinfo {author} {\bibfnamefont {C.~J.}\ \bibnamefont
  {Knight}}\ and\ \bibinfo {author} {\bibfnamefont {J.~S.}\ \bibnamefont
  {Hub}},\ }\href@noop {} {\bibfield  {journal} {\bibinfo  {journal} {Nucleic
  Acids Research}\ }\textbf {\bibinfo {volume} {43}},\ \bibinfo {pages}
  {W225--W230} (\bibinfo {year} {2015})}\BibitemShut {NoStop}%
\bibitem [{\citenamefont {Virtanen}\ \emph {et~al.}(2011)\citenamefont
  {Virtanen}, \citenamefont {Makowski}, \citenamefont {Sosnick},\ and\
  \citenamefont {Freed}}]{Virtanen2011}%
  \BibitemOpen
  \bibfield  {author} {\bibinfo {author} {\bibfnamefont {J.~J.}\ \bibnamefont
  {Virtanen}}, \bibinfo {author} {\bibfnamefont {L.}~\bibnamefont {Makowski}},
  \bibinfo {author} {\bibfnamefont {T.~R.}\ \bibnamefont {Sosnick}}, \ and\
  \bibinfo {author} {\bibfnamefont {K.~F.}\ \bibnamefont {Freed}},\ }\href@noop
  {} {\bibfield  {journal} {\bibinfo  {journal} {Biophysical Journal}\ }\textbf
  {\bibinfo {volume} {101}},\ \bibinfo {pages} {2061--2069} (\bibinfo {year}
  {2011})}\BibitemShut {NoStop}%
\bibitem [{\citenamefont {Minh}\ and\ \citenamefont
  {Makowski}(2013)}]{Minh2013}%
  \BibitemOpen
  \bibfield  {author} {\bibinfo {author} {\bibfnamefont {D.~D.~L.}\
  \bibnamefont {Minh}}\ and\ \bibinfo {author} {\bibfnamefont {L.}~\bibnamefont
  {Makowski}},\ }\href@noop {} {\bibfield  {journal} {\bibinfo  {journal}
  {Biophysical Journal}\ }\textbf {\bibinfo {volume} {104}},\ \bibinfo {pages}
  {873--883} (\bibinfo {year} {2013})}\BibitemShut {NoStop}%
\bibitem [{\citenamefont {Nguyen}\ \emph {et~al.}(2014)\citenamefont {Nguyen},
  \citenamefont {Pabit}, \citenamefont {Meisburger}, \citenamefont {Pollack},\
  and\ \citenamefont {Case}}]{Nguyen2014}%
  \BibitemOpen
  \bibfield  {author} {\bibinfo {author} {\bibfnamefont {H.~T.}\ \bibnamefont
  {Nguyen}}, \bibinfo {author} {\bibfnamefont {S.~A.}\ \bibnamefont {Pabit}},
  \bibinfo {author} {\bibfnamefont {S.~P.}\ \bibnamefont {Meisburger}},
  \bibinfo {author} {\bibfnamefont {L.}~\bibnamefont {Pollack}}, \ and\
  \bibinfo {author} {\bibfnamefont {D.~A.}\ \bibnamefont {Case}},\ }\href@noop
  {} {\bibfield  {journal} {\bibinfo  {journal} {The Journal of Chemical
  Physics}\ }\textbf {\bibinfo {volume} {141}},\ \bibinfo {pages} {22D508}
  (\bibinfo {year} {2014})}\BibitemShut {NoStop}%
\bibitem [{\citenamefont {Cammarata}\ \emph {et~al.}(2008)\citenamefont
  {Cammarata}, \citenamefont {Levantino}, \citenamefont {Schotte},
  \citenamefont {Anfinrud}, \citenamefont {Ewald}, \citenamefont {Choi},
  \citenamefont {Cupane}, \citenamefont {Wulff},\ and\ \citenamefont
  {Ihee}}]{Cammarata2008}%
  \BibitemOpen
  \bibfield  {author} {\bibinfo {author} {\bibfnamefont {M.}~\bibnamefont
  {Cammarata}}, \bibinfo {author} {\bibfnamefont {M.}~\bibnamefont
  {Levantino}}, \bibinfo {author} {\bibfnamefont {F.}~\bibnamefont {Schotte}},
  \bibinfo {author} {\bibfnamefont {P.~a.}\ \bibnamefont {Anfinrud}}, \bibinfo
  {author} {\bibfnamefont {F.}~\bibnamefont {Ewald}}, \bibinfo {author}
  {\bibfnamefont {J.}~\bibnamefont {Choi}}, \bibinfo {author} {\bibfnamefont
  {A.}~\bibnamefont {Cupane}}, \bibinfo {author} {\bibfnamefont
  {M.}~\bibnamefont {Wulff}}, \ and\ \bibinfo {author} {\bibfnamefont
  {H.}~\bibnamefont {Ihee}},\ }\href@noop {} {\bibfield  {journal} {\bibinfo
  {journal} {Nature methods}\ }\textbf {\bibinfo {volume} {5}},\ \bibinfo
  {pages} {881--886} (\bibinfo {year} {2008})}\BibitemShut {NoStop}%
\bibitem [{\citenamefont {Arnlund}\ \emph {et~al.}(2014)\citenamefont
  {Arnlund}, \citenamefont {Johansson}, \citenamefont {Wickstrand},
  \citenamefont {Barty},\ and\ \citenamefont {Williams}}]{Arnlund2014}%
  \BibitemOpen
  \bibfield  {author} {\bibinfo {author} {\bibfnamefont {D.}~\bibnamefont
  {Arnlund}}, \bibinfo {author} {\bibfnamefont {L.~C.}\ \bibnamefont
  {Johansson}}, \bibinfo {author} {\bibfnamefont {C.}~\bibnamefont
  {Wickstrand}}, \bibinfo {author} {\bibfnamefont {A.}~\bibnamefont {Barty}}, \
  and\ \bibinfo {author} {\bibfnamefont {G.~J.}\ \bibnamefont {Williams}},\
  }\href@noop {} {\bibfield  {journal} {\bibinfo  {journal} {Nature Methods}\
  }\textbf {\bibinfo {volume} {11}} (\bibinfo {year} {2014})}\BibitemShut
  {NoStop}%
\bibitem [{\citenamefont {Li}\ \emph {et~al.}(2011)\citenamefont {Li},
  \citenamefont {Pabit}, \citenamefont {Meisburger},\ and\ \citenamefont
  {Pollack}}]{Li2011}%
  \BibitemOpen
  \bibfield  {author} {\bibinfo {author} {\bibfnamefont {L.}~\bibnamefont
  {Li}}, \bibinfo {author} {\bibfnamefont {S.~A.}\ \bibnamefont {Pabit}},
  \bibinfo {author} {\bibfnamefont {S.~P.}\ \bibnamefont {Meisburger}}, \ and\
  \bibinfo {author} {\bibfnamefont {L.}~\bibnamefont {Pollack}},\ }\href
  {\doibase 10.1103/PhysRevLett.106.108101} {\bibfield  {journal} {\bibinfo
  {journal} {Physical Review Letters}\ }\textbf {\bibinfo {volume} {106}},\
  \bibinfo {pages} {108101} (\bibinfo {year} {2011})}\BibitemShut {NoStop}%
\bibitem [{\citenamefont {Tolokh}\ \emph {et~al.}(2014)\citenamefont {Tolokh},
  \citenamefont {Pabit}, \citenamefont {Katz}, \citenamefont {Chen},
  \citenamefont {Drozdetski}, \citenamefont {Baker}, \citenamefont {Pollack},\
  and\ \citenamefont {Onufriev}}]{Tolokh2014}%
  \BibitemOpen
  \bibfield  {author} {\bibinfo {author} {\bibfnamefont {I.~S.}\ \bibnamefont
  {Tolokh}}, \bibinfo {author} {\bibfnamefont {S.~A.}\ \bibnamefont {Pabit}},
  \bibinfo {author} {\bibfnamefont {A.~M.}\ \bibnamefont {Katz}}, \bibinfo
  {author} {\bibfnamefont {Y.}~\bibnamefont {Chen}}, \bibinfo {author}
  {\bibfnamefont {A.}~\bibnamefont {Drozdetski}}, \bibinfo {author}
  {\bibfnamefont {N.}~\bibnamefont {Baker}}, \bibinfo {author} {\bibfnamefont
  {L.}~\bibnamefont {Pollack}}, \ and\ \bibinfo {author} {\bibfnamefont
  {A.~V.}\ \bibnamefont {Onufriev}},\ }\href@noop {} {\bibfield  {journal}
  {\bibinfo  {journal} {Nucleic acids research}\ }\textbf {\bibinfo {volume}
  {42}},\ \bibinfo {pages} {10823--10831} (\bibinfo {year} {2014})}\BibitemShut
  {NoStop}%
\bibitem [{\citenamefont {Orthaber}, \citenamefont {Bergmann},\ and\
  \citenamefont {Glatter}(2000)}]{Orthaber2000}%
  \BibitemOpen
  \bibfield  {author} {\bibinfo {author} {\bibfnamefont {D.}~\bibnamefont
  {Orthaber}}, \bibinfo {author} {\bibfnamefont {A.}~\bibnamefont {Bergmann}},
  \ and\ \bibinfo {author} {\bibfnamefont {O.}~\bibnamefont {Glatter}},\
  }\href@noop {} {\bibfield  {journal} {\bibinfo  {journal} {Journal of Applied
  Crystallography}\ }\textbf {\bibinfo {volume} {33}},\ \bibinfo {pages}
  {218--225} (\bibinfo {year} {2000})}\BibitemShut {NoStop}%
\bibitem [{\citenamefont {Macke}\ and\ \citenamefont {Case}(1998)}]{Macke1998}%
  \BibitemOpen
  \bibfield  {author} {\bibinfo {author} {\bibfnamefont {T.}~\bibnamefont
  {Macke}}\ and\ \bibinfo {author} {\bibfnamefont {D.~A.}\ \bibnamefont
  {Case}},\ }\bibfield  {title} {\enquote {\bibinfo {title} {{Molecular Unusual
  Nucleic Acid Structures}},}\ }in\ \href@noop {} {\emph {\bibinfo {booktitle}
  {Molecular Modeling of Nucleic Acids}}},\ \bibinfo {editor} {edited by\
  \bibinfo {editor} {\bibfnamefont {N.~B.}\ \bibnamefont {Leontes}}\ and\
  \bibinfo {editor} {\bibfnamefont {J.}~\bibnamefont {{SantaLucia Jr.}}}}\
  (\bibinfo  {publisher} {American Chemical Society},\ \bibinfo {address}
  {Washington DC},\ \bibinfo {year} {1998})\ pp.\ \bibinfo {pages}
  {379--393}\BibitemShut {NoStop}%
\bibitem [{\citenamefont {Svergun}, \citenamefont {Barberato},\ and\
  \citenamefont {Koch}(1995)}]{Svergun1995}%
  \BibitemOpen
  \bibfield  {author} {\bibinfo {author} {\bibfnamefont {D.}~\bibnamefont
  {Svergun}}, \bibinfo {author} {\bibfnamefont {C.}~\bibnamefont {Barberato}},
  \ and\ \bibinfo {author} {\bibfnamefont {M.~H.}\ \bibnamefont {Koch}},\
  }\href@noop {} {\bibfield  {journal} {\bibinfo  {journal} {Journal of Applied
  Crystallography}\ }\textbf {\bibinfo {volume} {28}},\ \bibinfo {pages}
  {768--773} (\bibinfo {year} {1995})}\BibitemShut {NoStop}%
\bibitem [{\citenamefont {Case}\ \emph {et~al.}(2005)\citenamefont {Case},
  \citenamefont {Cheatham}, \citenamefont {Darden}, \citenamefont {Gohlke},
  \citenamefont {Luo}, \citenamefont {Merz}, \citenamefont {Onufriev},
  \citenamefont {Simmerling}, \citenamefont {Wang},\ and\ \citenamefont
  {Woods}}]{Case2005}%
  \BibitemOpen
  \bibfield  {author} {\bibinfo {author} {\bibfnamefont {D.~A.}\ \bibnamefont
  {Case}}, \bibinfo {author} {\bibfnamefont {T.~E.}\ \bibnamefont {Cheatham}},
  \bibinfo {author} {\bibfnamefont {T.}~\bibnamefont {Darden}}, \bibinfo
  {author} {\bibfnamefont {H.}~\bibnamefont {Gohlke}}, \bibinfo {author}
  {\bibfnamefont {R.}~\bibnamefont {Luo}}, \bibinfo {author} {\bibfnamefont
  {K.~M.}\ \bibnamefont {Merz}}, \bibinfo {author} {\bibfnamefont
  {A.}~\bibnamefont {Onufriev}}, \bibinfo {author} {\bibfnamefont
  {C.}~\bibnamefont {Simmerling}}, \bibinfo {author} {\bibfnamefont
  {B.}~\bibnamefont {Wang}}, \ and\ \bibinfo {author} {\bibfnamefont {R.~J.}\
  \bibnamefont {Woods}},\ }\href@noop {} {\bibfield  {journal} {\bibinfo
  {journal} {Journal of computational chemistry}\ }\textbf {\bibinfo {volume}
  {26}},\ \bibinfo {pages} {1668--88} (\bibinfo {year} {2005})}\BibitemShut
  {NoStop}%
\bibitem [{\citenamefont {Cheatham}, \citenamefont {Cieplak},\ and\
  \citenamefont {Kollman}(1999)}]{Cheatham1999}%
  \BibitemOpen
  \bibfield  {author} {\bibinfo {author} {\bibfnamefont {T.~E.}\ \bibnamefont
  {Cheatham}}, \bibinfo {author} {\bibfnamefont {P.}~\bibnamefont {Cieplak}}, \
  and\ \bibinfo {author} {\bibfnamefont {P.~A.}\ \bibnamefont {Kollman}},\
  }\href@noop {} {\bibfield  {journal} {\bibinfo  {journal} {Journal of
  biomolecular structure \& dynamics}\ }\textbf {\bibinfo {volume} {16}},\
  \bibinfo {pages} {845--862} (\bibinfo {year} {1999})}\BibitemShut {NoStop}%
\bibitem [{\citenamefont {P\'{e}rez}\ \emph {et~al.}(2007)\citenamefont
  {P\'{e}rez}, \citenamefont {March\'{a}n}, \citenamefont {Svozil},
  \citenamefont {Sponer}, \citenamefont {Cheatham}, \citenamefont {Laughton},\
  and\ \citenamefont {Orozco}}]{Perez2007}%
  \BibitemOpen
  \bibfield  {author} {\bibinfo {author} {\bibfnamefont {A.}~\bibnamefont
  {P\'{e}rez}}, \bibinfo {author} {\bibfnamefont {I.}~\bibnamefont
  {March\'{a}n}}, \bibinfo {author} {\bibfnamefont {D.}~\bibnamefont {Svozil}},
  \bibinfo {author} {\bibfnamefont {J.}~\bibnamefont {Sponer}}, \bibinfo
  {author} {\bibfnamefont {T.~E.}\ \bibnamefont {Cheatham}}, \bibinfo {author}
  {\bibfnamefont {C.~A.}\ \bibnamefont {Laughton}}, \ and\ \bibinfo {author}
  {\bibfnamefont {M.}~\bibnamefont {Orozco}},\ }\href@noop {} {\bibfield
  {journal} {\bibinfo  {journal} {Biophysical journal}\ }\textbf {\bibinfo
  {volume} {92}},\ \bibinfo {pages} {3817--29} (\bibinfo {year}
  {2007})}\BibitemShut {NoStop}%
\bibitem [{\citenamefont {Aqvist}(1990)}]{Aqvist1990}%
  \BibitemOpen
  \bibfield  {author} {\bibinfo {author} {\bibfnamefont {J.}~\bibnamefont
  {Aqvist}},\ }\href@noop {} {\bibfield  {journal} {\bibinfo  {journal}
  {Journal of physical chemistry}\ }\textbf {\bibinfo {volume} {94}},\ \bibinfo
  {pages} {8021--8024} (\bibinfo {year} {1990})}\BibitemShut {NoStop}%
\bibitem [{\citenamefont {Smith}\ and\ \citenamefont {Dang}(1994)}]{Smith1994}%
  \BibitemOpen
  \bibfield  {author} {\bibinfo {author} {\bibfnamefont {D.~E.}\ \bibnamefont
  {Smith}}\ and\ \bibinfo {author} {\bibfnamefont {L.~X.}\ \bibnamefont
  {Dang}},\ }\href@noop {} {\bibfield  {journal} {\bibinfo  {journal} {Journal
  of chemical physics}\ }\textbf {\bibinfo {volume} {100}},\ \bibinfo {pages}
  {3757--3766} (\bibinfo {year} {1994})}\BibitemShut {NoStop}%
\bibitem [{\citenamefont {Jorgensen}\ \emph {et~al.}(1983)\citenamefont
  {Jorgensen}, \citenamefont {Chandrasekhar}, \citenamefont {Madura},
  \citenamefont {Impey},\ and\ \citenamefont {Klein}}]{Jorgensen1983}%
  \BibitemOpen
  \bibfield  {author} {\bibinfo {author} {\bibfnamefont {W.~L.}\ \bibnamefont
  {Jorgensen}}, \bibinfo {author} {\bibfnamefont {J.}~\bibnamefont
  {Chandrasekhar}}, \bibinfo {author} {\bibfnamefont {J.~D.}\ \bibnamefont
  {Madura}}, \bibinfo {author} {\bibfnamefont {R.~W.}\ \bibnamefont {Impey}}, \
  and\ \bibinfo {author} {\bibfnamefont {M.~L.}\ \bibnamefont {Klein}},\
  }\href@noop {} {\bibfield  {journal} {\bibinfo  {journal} {Journal of
  chemical physics}\ }\textbf {\bibinfo {volume} {79}},\ \bibinfo {pages}
  {926–--935} (\bibinfo {year} {1983})}\BibitemShut {NoStop}%
\bibitem [{\citenamefont {Cheatham}\ and\ \citenamefont
  {Kollman}(1997)}]{Cheatham1997}%
  \BibitemOpen
  \bibfield  {author} {\bibinfo {author} {\bibfnamefont {T.~E.}\ \bibnamefont
  {Cheatham}}\ and\ \bibinfo {author} {\bibfnamefont {P.}~\bibnamefont
  {Kollman}},\ }\href@noop {} {\bibfield  {journal} {\bibinfo  {journal}
  {Structure}\ }\textbf {\bibinfo {volume} {5}},\ \bibinfo {pages} {1297--1311}
  (\bibinfo {year} {1997})}\BibitemShut {NoStop}%
\bibitem [{\citenamefont {Chen}\ and\ \citenamefont {Hub}(2014)}]{Chen2014}%
  \BibitemOpen
  \bibfield  {author} {\bibinfo {author} {\bibfnamefont {P.-C.}\ \bibnamefont
  {Chen}}\ and\ \bibinfo {author} {\bibfnamefont {J.~S.}\ \bibnamefont {Hub}},\
  }\href@noop {} {\bibfield  {journal} {\bibinfo  {journal} {Biophysical
  Journal}\ }\textbf {\bibinfo {volume} {107}},\ \bibinfo {pages} {435--447}
  (\bibinfo {year} {2014})}\BibitemShut {NoStop}%
\bibitem [{\citenamefont {Widom}\ and\ \citenamefont
  {Baldwin}(1980)}]{Widom1980}%
  \BibitemOpen
  \bibfield  {author} {\bibinfo {author} {\bibfnamefont {J.}~\bibnamefont
  {Widom}}\ and\ \bibinfo {author} {\bibfnamefont {R.~L.}\ \bibnamefont
  {Baldwin}},\ }\href@noop {} {\bibfield  {journal} {\bibinfo  {journal}
  {Journal of molecular biology}\ }\textbf {\bibinfo {volume} {144}},\ \bibinfo
  {pages} {431--453} (\bibinfo {year} {1980})}\BibitemShut {NoStop}%
\bibitem [{\citenamefont {Pelta}, \citenamefont {Livolant},\ and\ \citenamefont
  {Sikorav}(1996)}]{Pelta1996}%
  \BibitemOpen
  \bibfield  {author} {\bibinfo {author} {\bibfnamefont {J.}~\bibnamefont
  {Pelta}}, \bibinfo {author} {\bibfnamefont {F.}~\bibnamefont {Livolant}}, \
  and\ \bibinfo {author} {\bibfnamefont {J.~L.}\ \bibnamefont {Sikorav}},\
  }\href@noop {} {\bibfield  {journal} {\bibinfo  {journal} {Journal of
  Biological Chemistry}\ }\textbf {\bibinfo {volume} {271}},\ \bibinfo {pages}
  {5656--5662} (\bibinfo {year} {1996})}\BibitemShut {NoStop}%
\bibitem [{\citenamefont {Bulheller}\ and\ \citenamefont
  {Hirst}(2009)}]{Bulheller2009}%
  \BibitemOpen
  \bibfield  {author} {\bibinfo {author} {\bibfnamefont {B.~M.}\ \bibnamefont
  {Bulheller}}\ and\ \bibinfo {author} {\bibfnamefont {J.~D.}\ \bibnamefont
  {Hirst}},\ }\href@noop {} {\bibfield  {journal} {\bibinfo  {journal}
  {Bioinformatics}\ }\textbf {\bibinfo {volume} {25}},\ \bibinfo {pages}
  {539--540} (\bibinfo {year} {2009})}\BibitemShut {NoStop}%
\bibitem [{\citenamefont {Fischetti}\ \emph {et~al.}(2004)\citenamefont
  {Fischetti}, \citenamefont {Rodi}, \citenamefont {Gore},\ and\ \citenamefont
  {Makowski}}]{Fischetti2004}%
  \BibitemOpen
  \bibfield  {author} {\bibinfo {author} {\bibfnamefont {R.~F.}\ \bibnamefont
  {Fischetti}}, \bibinfo {author} {\bibfnamefont {D.~J.}\ \bibnamefont {Rodi}},
  \bibinfo {author} {\bibfnamefont {D.~B.}\ \bibnamefont {Gore}}, \ and\
  \bibinfo {author} {\bibfnamefont {L.}~\bibnamefont {Makowski}},\ }\href@noop
  {} {\bibfield  {journal} {\bibinfo  {journal} {Chemistry and Biology}\
  }\textbf {\bibinfo {volume} {11}},\ \bibinfo {pages} {1431--1443} (\bibinfo
  {year} {2004})}\BibitemShut {NoStop}%
\bibitem [{\citenamefont {Arnott}\ and\ \citenamefont
  {Hukins}(1972)}]{Arnott1972}%
  \BibitemOpen
  \bibfield  {author} {\bibinfo {author} {\bibfnamefont {S.}~\bibnamefont
  {Arnott}}\ and\ \bibinfo {author} {\bibfnamefont {D.~W.~L.}\ \bibnamefont
  {Hukins}},\ }\href@noop {} {\bibfield  {journal} {\bibinfo  {journal}
  {Biochemical and biophysical research communications}\ }\textbf {\bibinfo
  {volume} {47}},\ \bibinfo {pages} {1504--1509} (\bibinfo {year}
  {1972})}\BibitemShut {NoStop}%
\bibitem [{\citenamefont {P\'{e}rez}\ \emph {et~al.}(2008)\citenamefont
  {P\'{e}rez}, \citenamefont {Lankas}, \citenamefont {Luque},\ and\
  \citenamefont {Orozco}}]{Perez2008}%
  \BibitemOpen
  \bibfield  {author} {\bibinfo {author} {\bibfnamefont {A.}~\bibnamefont
  {P\'{e}rez}}, \bibinfo {author} {\bibfnamefont {F.}~\bibnamefont {Lankas}},
  \bibinfo {author} {\bibfnamefont {F.~J.}\ \bibnamefont {Luque}}, \ and\
  \bibinfo {author} {\bibfnamefont {M.}~\bibnamefont {Orozco}},\ }\href@noop {}
  {\bibfield  {journal} {\bibinfo  {journal} {Nucleic Acids Research}\ }\textbf
  {\bibinfo {volume} {36}},\ \bibinfo {pages} {2379--2394} (\bibinfo {year}
  {2008})}\BibitemShut {NoStop}%
\bibitem [{\citenamefont {Chen}\ and\ \citenamefont {Hub}(2015)}]{Chen2015}%
  \BibitemOpen
  \bibfield  {author} {\bibinfo {author} {\bibfnamefont {P.-C.}\ \bibnamefont
  {Chen}}\ and\ \bibinfo {author} {\bibfnamefont {J.~S.}\ \bibnamefont {Hub}},\
  }\href@noop {} {\bibfield  {journal} {\bibinfo  {journal} {Biophysical
  Journal}\ }\textbf {\bibinfo {volume} {108}},\ \bibinfo {pages} {2573--2584}
  (\bibinfo {year} {2015})}\BibitemShut {NoStop}%
\end{thebibliography}%

\end{document}